\documentstyle[12pt]{article}
\title{\bf Disclination vortices in elastic media}
\author{ \vspace*{2mm}
{\bf M. Pudlak$^{\dagger}$ and V.A. Osipov$^{\ddagger}$} \\
$^{\dagger}$\small \it Institute of Experimental Physics,
\small \it Slovak Academy of Science,\\
\small \it Watsonova 47, 043 53 Kosice, Slovak Republic\\
$^{\ddagger}$\small \it Bogoliubov Laboratory of Theoretical Physics,\\
\small \it Joint Institute for Nuclear Research, 141980 Dubna, Russia}

\begin{document}
\date{}
\maketitle

\begin{abstract}
The vortex-like solutions are studied in the framework of the
gauge model of disclinations in elastic continuum. A complete set
of model equations with disclination driven dislocations taken
into account is considered. Within the linear approximation an
exact solution for a low-angle wedge disclination is found to be
independent from the coupling constants of the theory. As a
result, no additional dimensional characteristics (like the core
radius of the defect) are involved. The situation changes
drastically for $2\pi$ vortices, where two characteristic lengths,
$l_\phi$ and $l_W$, become of importance. The asymptotical
behaviour of the solutions for both singular and nonsingular
$2\pi$ vortices is studied. Forces between pairs of vortices are
calculated.
\end{abstract}
\vskip 1cm


\section{Introduction}

Topologically nontrivial objects arising in various physically
interesting systems are the subject of considerable current
interest. It will suffice to mention the 't Hooft-Polyakov
monopole in the non-Abelian Higgs model, instantons in quantum
chromodynamics, solitons in the Skyrme model, Nielsen-Olesen
magnetic vortices in the Abelian Higgs model, etc. (see, e.g., the
book\cite{rajaraman}). Notice that similar objects are known in
condensed matter physics as well. For instance, vortices in
liquids and liquid crystals, solitons in low-dimensional systems
(e.g., in magnetics, linear polymers, and organic molecules) as
well as the famous Abrikosov magnetic vortices in superconductors
are the matter of common knowledge. Mathematically, all these
objects appear in the framework of nonlinear models as partial
solutions of strongly nonlinear equations. As is well-known, there
are still no general methods to study such equations. This makes
the derivation of the solutions difficult. An important point is
also that all the solutions are topologically stable and belong to
nontrivial homotopic sectors.

It should be noted that elastic media also leave room for
topological defects known as dislocations and disclinations.
However, these defects are usually treated within the linear
theory of elasticity. In this case, all the information about
defects is incorporated via source terms in the equilibrium
condition. The sources are assumed to have a $\delta$-function
form multiplied by either the Burgers or the Frank vector,
respectively (see, e.g.,~\cite{landau}). There were some attempts
to invoke the nonlinear theory of elasticity for description of
dislocations (see details in~\cite{gairola}) which account for the
nonlinear relation between stresses and strains thus giving a
possibility to determine stress and strain fields near a
dislocation at large deformations. Recently~\cite{jpa91},
nonlinear problems in dislocation theory were studied in the
framework of the gauge model of dislocations and disclinations
proposed in~\cite{edelen,lagoud}. Unfortunately, all these
approaches~\cite{gairola,jpa91} are based on the perturbation
scheme. It is known, however, that topologically nontrivial
solutions cannot be found in the framework of any perturbation
scheme.

As has been shown in~\cite{pla92,konoplich,jpa93,pla94}, the gauge
model~\cite{edelen} admits exact vortex-like solutions for wedge
disclinations. This finding confirms the view of a disclination as
a vortex of elastic medium. Notice that though the correspondence
between dislocations and vortices has been known for many years
(see, e.g., review~\cite{marcin}) the explicit vortex-like
solutions for disclinations were obtained only within the
gauge-theory approach. It is interesting to note that the elastic
flux due to rotational defects was found to be completely
determined by the gauge vector fields associated with
disclinations. In continuum there are no restrictions on the value
of the flux which is in fact the Frank vector, $\bf\Omega$. We
will consider in the present paper two cases: low-angle vortices
with $|{\bf\Omega}|\ll 1$, and $2\pi$ vortices with $|{\bf\Omega}|
= 2\pi$.

Before proceeding, let us mention that disclinations are of
importance in various crystalline and non-crystalline materials
(see, e.g.,~\cite{vladimirov}). Among the current applications one
can name the new class of carbon materials: fullerenes and
nanotubes (see, e.g.,~\cite{ebbesen}). According to the Euler's
theorem, these microcrystals can only be formed by having a total
disclination of $4\pi$. According to the geometry of the hexagonal
network it means the presence of twelve $60^{\circ}$ disclinations
on the closed hexatic elastic surface. Notice that disclinations
in liquid crystals is one of the best-studied cases. In
particular, the known exact solution, hedgehog, has been obtained
within the continuum model of nematics~\cite{landau}.

It is interesting that a hedgehog-like solution was also found for
a point $4\pi$ disclination in the framework of the gauge
model~\cite{pla90,pa91}. An important advantage of the gauge model
follows from the fact that it is similar to the known field theory
models, first of all to the non-Abelian and Abelian Higgs models,
where topological objects are studied well. Taking into account
this similarity we have found two exact static solutions for
linear disclinations~\cite{pla92,jpa93,pla94} which will be
discussed in this paper. In particular, within the linear scheme
our model recovers the equilibrium conditions of the standard
elasticity theory with a disclination-induced source being
generated by gauge fields. It should be noticed, however, that
these solutions were obtained within the restricted model where
the dislocation-induced contribution to the Lagrangian was
neglected. As a matter of fact, this contribution always exists
(so-called disclination driven dislocations~\cite{edelen}). The
goal of the present paper is to consider the most general model
for linear disclinations in elastic continuum involving all main
contributions.

The paper is structured as follows. In section 2 the gauge model
of disclinations is presented. The Lagrangian describes elastic
deformations and self-energy of disclinations as well as a
contribution from the rotational dislocations. A complete set of
field equations is formulated for the static case. In section 3
the gauge model is studied within the linear approximation. An
exact vortex-like solution for a low-angle straight wedge
disclination is found to be independent from the coupling
constants of the theory. Forces between vortices as well as dipole
configurations are studied. Topologically stable $2\pi$ vortices
are considered in section 4 as the solutions of strongly nonlinear
equations of the theory. First, we consider an exact singular
solution and show that the dislocation-induced contribution
becomes important. Second, we analyze briefly a problem of the
nonsingular vortex. In particular, appropriate asymptotic
solutions are found. Section 5 is devoted to concluding comments.

\section{Gauge model}
\medskip

In accordance with the basic assumption of the gauge
approach~\cite{edelen}, topological defects can be introduced into
the Lagrangian of elasticity theory through gauge fields. Namely,
the defect-free Lagrangian of elasticity theory is invariant under
the homogeneous action of the space group $G=SO(3)\triangleright
T(3)$. Let us consider this group as the gauge one and assume that
the inhomogeneous action of $G$ leaves the initial Lagrangian
invariant. In this case, we arrive at the Yang-Mills-type theory
which contains two different gauge fields associated with
rotational $SO(3)$ and translational $T(3)$ group, respectively.
The model is based on the Yang-Mills minimal coupling arguments.
While the initial Lagrangian is chosen to be quadratic in
distortion fields, the presence of gauge fields makes it strongly
nonlinear\footnote{The cubic terms can be included into the
Lagrangian as well (see~\cite{lagoud}), but in so doing the model
becomes too cumbersome for analysis.}. That is why the nonlinear
relation between stresses and strains occurs from the very
beginning.

For disclinations and rotational dislocations, only $SO(3)$ group
should be taken into account. The problem becomes simpler for
rectilinear defects, which are of interest here. Indeed, in this
case the rotational symmetry becomes broken only in the plane
normal to the defect line, and, correspondingly, the gauge group
reduces to $SO(2)$. Notice that such gauge group is typical for
models containing vortex-like objects.

The Lagrangian that is invariant under the inhomogeneous action of
the $SO(2)$ gauge group takes the following form~\cite{jpa93}
\begin{equation}
{\cal L}={\cal L}_{\chi}+{\cal L}_{\phi}+{\cal L}_{W},
\end{equation}
where
\begin{equation}
{\cal L}_{\chi}=\frac{1}{2}\rho_{0}B_3^iB_3^i-\frac{1}{8}
[\lambda(\mbox{tr} E)^2+2\mu \mbox{tr} E^2]
\end{equation}
describes the elastic properties of the material while
\begin{equation}
{\cal L}_{\phi}=-\frac{1}{2}s_{1}D_{ab}^ik^{ac}k^{bd}D_{cd}^i
\end{equation}
and
\begin{equation}
{\cal L}_{W}=-\frac{1}{2}s_{2}F_{ab}g^{ac}g^{bd}F_{cd}
\end{equation}
describe defect-induced contributions. Here $E_{AB}=B_A^i
B_B^i-\delta_{AB}$ is the strain tensor,
$D_{ab}^i=\epsilon_j^iF_{ab}\chi^j$,
$F_{ab}=\partial_{a}W_{b}-\partial_{b}W_{a}$, $s_1$ and $s_2$ are
the coupling constants, tr$E = E_{AA}$ and summation over repeated
indices is assumed. In accordance with the minimal replacement
arguments, the distortion tensor is written as
\begin{equation}
B_a^i=\partial_a\chi^i+\epsilon_j^{i}\chi^jW_a
\end{equation}
where $\chi^i(x^a)=\chi^i(x^A,T)$ characterizes the configuration
at time T in terms of the coordinate cover $(x^A)$ of a reference
configuration, $W_a$ is the compensating gauge field associated
with the disclination field. In (3) the tensor $k^{ab}$ is given
by $k^{AB}=-\delta^{AB}$, $k^{33}=1/y$, and $k^{ab}=0$ for
$a\not=b$, whereas in (4) $g^{AB}=-\delta^{AB}$, $g^{33}=1/\xi$,
and $g^{ab}=0$ for $a\not=b$. The parameters $y$ and $\xi$ are the
two positive "propagation parameters", $\epsilon_j^{i}$ is a
completely antisymmetric tensor, $\epsilon_2^{1}=1$, and $\lambda$
and $\mu$ are the Lam\'{e} constants. (Indices a,b,c,...=1,2,3 and
A,B,C,...=1,2 are the space labels, whereas i,j,k,...=1,2 belong
to $SO(2)$).

Notice that ${\cal L}_W$ describes a self-energy of pure
rotational defects (disclinations). In accordance with the general
approach it acquires a standard form of the $SO(2)$ Yang-Mills
action. ${\cal L}_{\phi}$ is an additional invariant term
associated with translational defects (rotational dislocations).
This is reflected in the fact that the tensor $D^i_{ab}$ in (3) is
directly related to the density of dislocations defined as
$\alpha^{Ai}=\epsilon^{ABC}D^i_{BC}$ (see~\cite{edelen}). Thus the
appearance of this term in the Lagrangian reflects the known in
the defect theory fact that the presence of disclinations implies
the presence of dislocations.

The Euler-Lagrange equations for (1) take the following form in
the static case
\begin{equation}
\partial_A\sigma_k^A=\epsilon_k^jW_{C}\sigma_j^C+s_{1}\chi^l
\delta_{lk}F_{AB}F^{AB},
\end{equation}
\begin{equation}
\sigma_A^j\delta_{ij}\epsilon_l^i\chi^l=2\partial_B[(s_{1}(\chi)^2
+s_{2})F^{BA}],
\end{equation}
where $(\chi)^2=\chi^l\chi_{l}, \ l=1,2$. To avoid cumbersome
expressions we will sometimes omit the right order of the top and
bottom indices which can be easily restored by using the
appropriate $\delta$-symbols. The stress tensor $\sigma_C^j$ is
determined to be
\begin{equation}
\sigma_C^j=\frac{1}{2}[\lambda(\mbox{tr} E)B_C^j+ 2\mu(E_{CB}
B_B^j)].
\end{equation}

It is convenient to introduce the dimensionless variables via
$x^i=\sqrt{s_2/s_1}\tilde{x}^i$ and
$W_A=\sqrt{s_1/s_2}\tilde{W}_A$. The Euler-Lagrange equations (6)
and (7) become
\begin{equation}
\tilde{\partial}_A\tilde{\sigma}_k^A=\epsilon_k^j\tilde{W}_C
\tilde{\sigma}_j^C + \frac{s_1^2}{\mu s_2 }\tilde{\chi}^l
\delta_{lk}\tilde{F}_{AB}\tilde{F}_{AB},
\end{equation}
\begin{equation}
\tilde{\sigma}_A^j \delta_{ij} \epsilon_l^i \tilde{\chi}^l=
\frac{2s_1^2}{\mu s_2 } \tilde{\partial}_B[((\tilde{\chi})^2+1)
\tilde{F}^{BA}],
\end{equation}
where $\chi^l(x^B)=\sqrt{s_2/s_1}\tilde{\chi}^l
(\tilde{x}^B)$,
$\partial_B=\sqrt{s_1/s_2}\tilde{\partial}_B$,
$\tilde{F}_{AB}=\tilde{\partial}_A\tilde{W}_B-
\tilde{\partial}_B\tilde{W}_A$ and the
stress tensor is found to be
\begin{equation}
\tilde{\sigma}_j^A=\frac{L}{2}(\mbox{tr} \tilde{E}) \tilde{B}_A^j
+ \tilde{E}_{AC}\tilde{B}_C^j.
\end{equation}
Here $L=\lambda /\mu$, and the strain tensor takes the form
\begin{equation}
\tilde{E}_{AB}=\tilde{B}_A^i\tilde{B}_B^i-\delta_{AB}
\end{equation}
with $\tilde{B}_A^i=\tilde{\partial}_A\tilde{\chi}^i +
\epsilon_j^i\tilde{\chi}^j\tilde{W}_A$. To simplify notation,
below we will omit the "tilde" symbol. As is seen, the
self-consistent system of field equations (9) and (10) is strongly
nonlinear. In some respects (see discussion in~\cite{pla92,jpa93})
it is similar to that in the Abelian Higgs model. An analogous
system of equations appears under study of type-II superconductors
in magnetic field directed along the z-axis. This observation was
helpful in finding of the exact vortex-like solution to (9) and
(10) for $s_1=0$~\cite{pla92}. A possible way to study these
equations is the linearization procedure which is valid for
low-angle defects. It was found~\cite{pla94} that there is an
exact vortex-like solution at $s_1=0$ which reproduces the known
strain and stress fields of straight wedge disclinations. Notice
that both solutions are found to be singular at the disclination
line. While this singularity is well known in dislocation theory,
the question arises if there exist nonsingular solutions. Besides,
it turns out that for $s_1=0$ the coupling constant $s_2$ drops
out of the problem as well. Let us reexamine the vortex-like
solutions within the most general model.

\section{Low-angle disclination vortices}

Let us consider first the linearized equations of the model. The
linearization procedure is based on a homogeneous scaling of the
gauge group generators (see details in~\cite{edelen}). It is clear
that employing the linear approximation we restrict our
consideration to small deformations or, correspondingly, to the
low-angle defects. The displacement vector ${u}^i$ can be
introduced as follows
\begin{equation}
{\chi}^i({x}^B)=\delta_A^i{x}^A+{u}^i({x}^B).
\end{equation}
Then, with the scaling parameter $\epsilon$ all the fields are
expanded in series of $\epsilon$
\begin{equation}
{u}^i=\epsilon{u}_1^i+\epsilon^2{u}_2^i+...,\quad
{W}_A=\epsilon{W}_{1A}+\epsilon^2{W}_{2A}+...
\end{equation}
Taking into account only defect-induced displacements ${u}_1^i$
that are of interest here, we get the first order equations in the
following form
\begin{equation}
{\partial}_B[({x}^2+{y}^2+1){F}^{BA}]=0,
\end{equation}
\begin{equation}
{\Delta}{\bf{u}}+(L+1){\nabla}div{\bf{u}}={\bf j}+
{\bf{x}}\frac{s_1^2}{\mu s_2}{F}_{AB}{F}^{AB}
\end{equation}
where the density of elastic flow due to a disclination, $\bf{j}$,
is completely determined by gauge fields
\begin{equation}
j_{C}=-(L-1)\epsilon_{AC}W_A -
L\epsilon_{AB}x_B\partial_{C}W_A - \epsilon_{AB}x_B\partial_A
W_{C} - \epsilon_{C B}x_B\partial_A W_A.
\end{equation}
Hereafter we omit the index 1 denoting the order of the
approximation. Let us emphasize that we assume here $s_1^2/\mu s_2
\sim 1/\epsilon$. For other two possibilities, $s_1^2/\mu s_2 \sim
{\epsilon}$ and $s_1^2/\mu s_2 \sim 1$, terms with $s_1$ in (15)
and (16) are of little importance thus the standard
theory~\cite{pla94} is recovered. Notice that for $s_1=0$ a
solution of these equations was found in~\cite{pla94}. An
interesting property of this solution is its independence from the
parameter $s_2$ as well. This can be seen directly from (15) and
(16) where $s_2$ is completely absent. Let us choose the following
vortex-like ansatz
\begin{equation}
W_A = -W(r)\epsilon_{AB}\partial_B\log r,
\quad {u}^i={x}^i{G}({r})
\end{equation}
where ${r}^2={x}^2+{y}^2$. With (18) taken into account one can
rewrite (15) and (16) as follows
\begin{equation}
{\partial}_{{r}}\left [({r}^2+1)\frac{{W}'
({r})}{{r}}\right ]=0,
\end{equation}
\begin{equation}
(L+2)\left ({G}''({r})+\frac{3{G}'({r})} {{r}}\right ) =
L\frac{{W}'({r})}{{r}} -\frac{2{W}({r})}{{r}^2} +\frac{2s_1^2}{\mu
s_2}\left (\frac{{W}'({r})}{{r}}\right )^2
\end{equation}
where ${G}'$ stands for $d{G}/d{r}$ and ${W}'$ for $d{W}/d{r}$. A
solution to (19) takes the form
\begin{equation}
{W}({r})=C_1\ln({r}^2+1)+C_0
\end{equation}
with $C_0$ and $C_1$ being the integration constants. Notice that
a disclination flow through the plane xy is given by a circular
integral
\begin{equation}
\frac{1}{2\pi} \oint{\bf{W}}d{\bf{r}}=\nu.
\end{equation}
Taking this into account, we immediately obtain that $C_1=0$. Thus
the constant $C_0$ turns out to be in fact a topological
characteristic of the defect, that is the Frank index $\nu$. For
${W}({r})=\nu$ (20) becomes remarkably simpler and has a solution
\begin{equation}
{G}({r})=-\frac{\nu}{L+2}\ln{r} -\frac{1}{2}C_2{r}^{-2}+C_3.
\end{equation}
Since the boundary condition requires $u^i(0)=0$ we must put
$C_2=0$. Turning back to the dimensional variables, we finally
obtain
\begin{equation}
u^i=-x^i(\frac{\nu}{L+2}\ln\sqrt{\frac{s_1}{s_2}}r+C_3),
\end{equation}
where $C_3$ is still an arbitrary constant. As is seen, the term
with $s_1$ and $s_2$ only renormalizes the constant $C_3$. As an
example, for the straight wedge disclination on a disk of radius R
with a boundary condition in the form $u^i(R)=0$ one obtains
\begin{equation}
u^i=-x^i\frac{\nu}{L+2}\ln\frac{r}{R}.
\end{equation}
Similarly, for the most-used boundary condition,
$\sigma_{kl}n_l=0$ at the free surfaces, one can reproduce the
well known stress fields for a wedge disclination on a disk (see
details in~\cite{pla94}). It is seen that parameters $s_1$ and
$s_2$ actually drop out from (24) and (25). Thus one can conclude
that the information carried by the coupling constants $s_1$ and
$s_2$ is lost within the linear approximation. What does it mean?
As is known, the classical theory of elasticity introduces a
characteristic velocity $\sqrt{\mu/\rho_0}$, but does not lead to
a characteristic length. For this reason there is no room within
the linear theory of elasticity for description of the core
region. It is interesting that the gauge theory of
defects~\cite{edelen,lagoud} introduces appropriate length scales.
These are the dislocation length scale, $l_{\phi}^2=s_1/\mu$, and
the disclination length scale, $l_W^4=s_2/\mu$. Nevertheless, as
we have just seen, in the linear approximation the gauge theory
loses these parameters thus making the description of the core
region impossible. One can expect\footnote{We would like to thank
Prof. A.M. Kosevich for attracting our attention to this
possibility.}, however, that these parameters would be of
importance in a study of the basic model equations (6) and (7). We
will consider this problem in section 4.

\subsection{Forces between vortices, dipole configurations}

Let us consider two low-angle vortices with parallel Frank vectors
oriented along the z-axis. This corresponds to a pair of straight
wedge disclinations. For simplicity, we suppose that disclination
lines coincide with their axes of rotations. The stress field due
to the disclination results in the force acting on the second
defect (in perfect analogy to the known Peach-Koehler force in
dislocation theory). Generally it can be written as~\cite{das}
(per unit length of the disclination line)
\begin{equation}
{\cal
F}_{c}=\epsilon_{bkc}\epsilon_{amn}\Omega_mX_n\sigma_{ab}\xi_k,
\end{equation}
where $\vec\xi$ is a tangent vector at the disclination line,
$\Omega_m$ are the components of the Frank vector,
$X_n=x_n-x_n^0$, and $x_n^0$ is a point on the axis of the
disclination. In our case, $x_n^0=0$ and one has to put in (26)
$k=3$ and $m=3$. Notice that the same expression follows from the
general equations of the gauge model~\cite{edelen,lagoud} in the
linear approximation. As was shown above, the stress fields take
the well-known in the linear theory form. Evidently, the force
between two parallel low-angle wedge disclinations also has the
known form. For example, when the first vortex is situated at
point (0,0) while the second one at point (d,0) on the plane xy we
obtain from (26)
\begin{equation}
{\cal F}_x = d\Omega\sigma_{yy}, \qquad {\cal
F}_y=-d\Omega\sigma_{yx}
\end{equation}
in accordance with~\cite{das}. Notice that for vortices with equal
but oppositely directed Frank vectors such configuration
corresponds to a wedge disclination dipole with nonskew axes of
rotations. It is interesting to reproduce the solution for the
dipole within the gauge model. Since the previous analysis shows
that the constants $s_1$ and $s_2$ are inessential in the linear
approximation, we will drop terms with $s_1$ and put $W(r)=\nu$
from the beginning. A dipole solution to (15) then reads
\begin{equation}
W_B = -\nu\epsilon_{BC}\partial_C(\log r_1 - \log r_2).
\label{eq:4}
\end{equation}
The most simple way to solve (16) is via the Airy stress function.
Namely, let us differentiate (16). After straightforward
calculations one can rewrite this equation as
\begin{equation}
(\lambda/4\mu+1/2)\,\Delta\,\mbox{tr} E = \epsilon_{A
B}\partial_A W_B.
\label{eq:5}
\end{equation}
The last term in the right-hand side of (\ref{eq:5}) describes a
source due to disclination fields. For solution (\ref{eq:4}) it
takes the form $$\epsilon_{A B}\partial_A W_B=
\nu\Delta(\log r_1 - \log r_2)=2\pi\nu(\delta (\vec r_1)-\delta
(\vec r_2)).$$ Introducing the Airy stress function $\chi(\vec r)$
by $\sigma_{BA}=\epsilon_{BM}\epsilon_{A N}\partial_M
\partial_N\chi(\vec r)$,
one can finally rewrite the equation (\ref{eq:5}) as
\begin{equation}
K_0^{-1}\Delta^2\chi = 2\pi\nu(\delta (\vec r_1)-\delta (\vec
r_2)). \label{eq:6}\end{equation} Here $K_0=4\mu(\lambda
+\mu)/(\lambda +2\mu)$, and tr$E=(1/(\lambda+\mu))\Delta\chi(\vec
r)$. Evidently, a solution to (\ref{eq:6}) is the sum
$\chi=\chi_1+\chi_2$ with $\chi_i=Ar^2_i\ln r_i$ (i=1,2). One can
easily find that $A=\pm\nu K_0/4$. Finally, turning back to
$\sigma_{BA}$ one can exactly reproduce the known stress
fields for disclination dipoles (cf.,
e.g.,~\cite{vladimirov,wit}).

\section{2$\pi$ disclination vortices}

Let us choose the following ansatz for (6) and (7) to meet the
necessary symmetry requirements
\begin{equation}
\chi^1({x}^A)=F({r})\cos\theta, \quad
\chi^2({x}^A)=F({r})\sin\theta
\end{equation}
and
\begin{equation}
{W}_x({x}^A)=-\frac{{y}}{{r}^2}W({r}),\quad
{W}_y({x}^A)=\frac{{x}}{{r}^2} W({r})
\end{equation}
where ${r}^2={x}^A{x}_A$ (${r},\theta$ are the polar coordinates).
All the variables here are again dimensionless. We restrict
ourselves by the topological sector with $n=1$. As is seen, (31)
and (32) describe a $2\pi$ vortex, that is the circular integral
in (22) is equal to $\nu=1$. With (31) and (32) taken into account
(9) and (10) reduce to
\[4\frac{F}{{r}^2}W'^2=\frac{W-1}{{r}}f
\left[ K(g^2+f^2-2)+2P(f^2-1) \right]+ K\left[\frac{d}{d{r}}(f^2
g)+ \frac{f^2g}{{r}}\right]\]
\begin{equation}
+\left[3(K+2P)g'g^2-2(K+P)g'\right]
+\frac{1}{{r}}\left[(K+2P)g^3-2(K+P)g\right],
\end{equation}
\begin{equation}
4\frac{d}{d{r}}\left[(1+F^2)
\frac{W'}{{r}}\right]=Ff[K(f^2+g^2-2)+2P(f^2-1)]
\end{equation}
where $K=\lambda s_2/s_1^2$, $P=\mu s_2/s_1^2$, $g=dF(r)/dr$,
$f=F(r)(1-W(r))/r$, $W'=dW(r)/dr$. This system of equations is of
our interest in this section. We will consider two possible cases.

\subsection{Singular vortex}

In the case of $s_1=0$ an exact solution to (33) and (34) for a
static disclination vortex was found in~\cite{pla92, konoplich}.
The solution is singular on the disclination line with $W(r)=1$.
It is interesting to note that the same solution is valid for the
general case when $s_1\neq 0$. Indeed, for $W(r)=1$ one gets
$f=0$. In such an event, both sides of (34) turn out to be zero
whereas (33) reduces to
\begin{equation}
\left(3g'g^2-N_0^2g'\right) +\frac{1}{r}\left(g^3-N_0^2g\right)=0
\end{equation}
where $N_0^2=2(K+P)/(K+2P)=2(\lambda+\mu)/(\lambda+2\mu)$.
Carrying out an integration in (35) one obtains finally the
algebraic equation
\begin{equation}\label{eq32}
  |g^3(r) - N_0^2g(r)| = \frac{C_0}{r}
\end{equation}
where $C_0$ is an arbitrary integration constant. The solution to
(36) is written as $g(r)=(2/\sqrt{3})N_0\bar g(r)$ with
\begin{equation}\label{eq33}
\bar g(r)={\left\{ {\bar
g_1(r)=\cosh[\frac{1}{3}\cosh^{-1}(r_0/r)], \qquad \ r\leq r_0}
\atop {\bar g_2(r)=-\cos[\frac{1}{3}\cos^{-1}(r_0/r)+\frac{2\pi
l}{3}], \quad r\geq r_0}\right.,}
\end{equation}
where $r_0=3{\sqrt 3}C_0/2N_0^3$ is a characteristic parameter,
and $l=0,1,2$. Notice that in the natural variables $r_0$ becomes
dimensional. It was supposed in~\cite{pla92} that $r_0$ could be
considered as a core radius of the defect. Indeed, according to
(37) the point $r=r_0$ is prominent. It should be mentioned,
however, that this attractive possibility for description of the
core radius requires an involvement of the additional
phenomenological parameter, $C_0$, into the theory. At the same
time, the model parameters $s_1$ and $s_2$ drop out of (35) and
(37).

The main reason is that the chosen ansatz $W(r)=1$ (the pure gauge
for all $r$) is too restrictive. Obviously, any solutions with no
constant $W(r)$ are of special interest. However, equations (33)
and (34) look rather cumbersome, and a search for nontrivial exact
solutions remains still an open problem. Let us try another way of
looking at the problem. For this purpose, one can put $C_0=0$ in
(37). The exact solution to (35) takes then the essentially
simpler form
\begin{equation}
W(r)=1, \qquad F_{1,2}(r)=\pm N_0r, \qquad F_3(r)=0.
\end{equation}
As a first step, let us consider small perturbations of the exact
solution (38). It will be shown below that even this simplified
consideration allows us to get an important information about the
role of coupling constants $s_1$ and $s_2$ in the theory. Let us
write
\begin{equation}
F(r)=N_0r +\epsilon u_1(r)+\epsilon^2 u_2(r)+.....
\end{equation}
\begin{equation}
W(r)=1-\epsilon w_1(r)+.....
\end{equation}
We consider here the case $F(r)=F_1(r)$. Substituting this
expansion into (33) and (34) one gets
\begin{equation}
\left(\frac{1}{r}+N_0^2r\right)
w_1''(r)+\left(N_0^2-\frac{1}{r^2}\right)w_1'(r)-
\frac{P}{2}N_0^4rw_1(r)=0,
\end{equation}
\begin{equation}
u_2''(r)+\frac{1}{r}u_2'(r)= \frac{N_0}{(K+P)}\frac{(w_1')^2}{r}
-N_0\frac{w_1^2}{2r}-\frac{KN_0}{K+2P} w_1w_1'.
\end{equation}
It should be noted that $u_1=0$. This follows from the requirement
$F(r)\rightarrow 0$ at $r\rightarrow 0$. Let us analyze (41). As
is known, the equation of the type $w''+H(x)w'+Q(x)w=0$ can be put
into the form $z''+I(x)z=0$ by a substitution
$$w(x)=z(x)\exp\left(-\frac{1}{2}\int H(x)dx\right)$$ where
$I=-\frac{1}{2}H'-\frac{1}{4}H^2+Q$. For (41)
\begin{equation}
H(r)=\frac{N_0^2r^2-1} {r\left(N_0^2r^2+1\right)},\qquad
Q(r)=-\frac{PN_0^4r^2}{2(N_0^2r^2+1)}
\end{equation}
In this case, the general solution is not known yet. Instead, let
us derive two limiting cases.

1.In the limit $N_0^2r^2\gg 1$ one obtains $H(r)=1/r$, and
$w_1(r)=z(r)/\sqrt{r}$. The equation for $z(r)$ takes the form
\begin{equation}
z''(r) = \left(\frac{PN_0^2}{2} - \frac{1+2P}{4r^2}\right)z(r).
\end{equation}
This is a special form of the Whittaker equation with the solution
\begin{equation}
z(r)=C_lW_{0,m}(\beta r)
\end{equation}
where $C_l$ is a constant, $W_{0,m}$ is the Whittaker function,
$m=\pm i\sqrt{P/2}$, and $\beta =N_0\sqrt{2P}$. In this case,
$W(r)$ is found to be
\begin{equation}
W(r)=1-C_lr^{-1/2}W_{0,m}(\beta r).
\end{equation}
Notice that $C_l$ includes $\epsilon$. Depending on $\beta$ two
asymptotics can be obtained.

(i) For $\beta r\gg 1$ one gets
\begin{equation}
W(r)=1-C_lr^{-1/2}\exp(-\frac{1}{2}\beta r)
\end{equation}
and
\begin{equation}
F(r)=N_0r+C_l^2\frac{K}{4(K+2P)} \frac{1}{N_0r}\exp(-\beta r).
\end{equation}

(ii) In the limit $\beta r\ll 1$ one has
\begin{equation}
W(r)=1-C_l\cos\left (\sqrt{P/2}\ln(\beta r)\right )
\end{equation}
and
\begin{equation}
F(r)=\frac{N_0}{2}(2-C_l^2)r.
\end{equation}

2. Let us consider the limit $N_0^2r^2\ll 1$. In this case,
$H(r)=-1/r$ and $w_1(r)= \sqrt{r}z(r)$. The equation for $z(r)$
reads
\begin{equation}
z''(r) = \left(\frac{3}{4r^2}+\frac{P}{2}N_0^4r^2\right)z(r)
\end{equation}
with a solution in the form
\begin{equation}
z(r)=2C_s\frac{\sinh(\frac{1}{2}\gamma^2r^2)}{\sqrt{\gamma r}}
\end{equation}
where $\gamma^2=N_0^2\sqrt{P/8}$. Thus,
\begin{equation}
W(r)=1-\frac{2C_s}{\sqrt{\gamma}}\sinh(\frac{1}{2} \gamma^2r^2).
\end{equation}
Two limiting cases are of interest.

(i) For $\gamma^2 r^2\gg 1$ one obtains
\begin{equation}
W(r)=1-\frac{C_s}{\sqrt{\gamma}}\exp(\frac{1}{2} \gamma^2r^2)
\end{equation}
and
\begin{equation}
F(r)=N_0r+\Omega_1C_s^2r^3\sum_{n=0}
^{\infty}\frac{(\gamma^2r^2)^n}{n!}\frac{1}{(2n+3)^2}
\end{equation}
where $\Omega_1=N_0\gamma(\sqrt{2P}-K)/(K+2P)$.

(ii) If $\gamma^2 r^2\ll 1$ we get
\begin{equation}
W(r)=1-C_s\gamma^{3/2}r^2
\end{equation}
and
\begin{equation}
F(r)=N_0r+\frac{4}{9}C_s^2\gamma^3\frac{N_0}{K+P}r^3.
\end{equation}
For better understanding of the obtained results let us return to
the dimensional variables. It is interesting that both
characteristic lengths of the theory turn out to be involved.
Namely, the dimensional $r$ reads $r=\sqrt{s_2/s_1}\tilde r$, and
the important parameters $\beta$ and $\gamma$ are determined via
$\sqrt{P}=l_W^2/l_{\phi}^2$. In~\cite{lagoud} three physically
interesting limits were discussed. In particular, in accordance
with~\cite{lagoud} the typical condition which is valid in
crystals and polycrystals is $l_W\gg l_{\phi}$, in some
polycrystals and amorphous bodies $l_W\sim l_{\phi}$ while the
most exotic limit which can be expected in some special amorphous
materials is $l_W\ll l_{\phi}$. Our consideration shows that
vortex-like solutions have different asymptotics in each case. For
$l_W\gg l_{\phi}, l_W\sim l_{\phi}$, and $l_W\ll l_{\phi}$ they
follow (1(i),2(i)), (1(i),2(ii)), and (1(ii),2(ii)), respectively
(see above). In accordance with (11) and (12) this results in
different asymptotic behaviour of strains and stresses due to
disclinations thus giving a possibility for the experimental
verification. On the other hand, the obtained results indicate
that the proper information about the core region, if any, can be
obtained only within the framework of the complete gauge model
which should include rotational dislocations.

\subsection{Nonsingular $2\pi$ vortex}

Let us discuss briefly a possibility for a nonsingular solution in
(31) and (32) which provides a finite energy of the vortex. This
means that condition $W(r)\rightarrow 0$ for $r\rightarrow 0$
should be satisfied. A simple analysis of (33) and (34) shows that
there is an asymptotic solution at small $r$ in the form
\begin{equation}
W(r)\sim r^{\alpha}, \qquad F(r)\sim ar^{\mu},
\end{equation}
where $\mu=1$, $\alpha=2$, and $a$ is an arbitrary constant with
the only restriction $a\neq 1$ following from (34). Notice that
this resembles the behaviour of the Abrikosov-Nielsen-Olesen
vortex. For large $r$, the asymptotics found in the previous
subsection are valid. To prove the existence of the solution for
any $r$ the numerical calculations of variations in the energy
density which has the following form
\begin{eqnarray}
E(r) &=& \lambda\left[F'^2+\frac{F^2}{r^2}(1-W)^2-2\right]^2
+2\mu\left[F'^4+\frac{F^4}{r^4}(1-W)^2-2F'^2-2\frac{F^2}{r^2}
(1-W)^2+2\right] \nonumber \\
&+&\frac{s_1^2}{s_2}\left[(F^2+1)\left(\frac{W'}{r}\right)^2\right]
\end{eqnarray}
where $F'=dF/dr$ and $W'=dW/dr$ must be used. This study will give
a final answer about whether the above asymptotics come from the
unique solution or not. The corresponding calculations are now in
progress.

\subsection{Forces}
Let us discuss briefly the force between two $2\pi$ vortices. In
accordance with the classical formula (26) one obtains
\begin{equation}
{\cal F}_{A}=\Omega f\left[\lambda (g^{2}+f^{2}-2)+2\mu (f^{2}-1)
\right]x_A.
\end{equation}
Thus the force turns out to be exactly zero for the solution
$W(r)=1$,$F(r)=N_{0}r$, that is in the case of $2\pi$ singular
vortices. Assuming a small perturbation of this exact solution we
get the following asymptotics:\\
1. For
$N_{0}^{2}r^{2}s_1/s_2\gg1$
\begin{equation}
{\cal F}_{A}=-2\mu
N_{0}^{2}C_{l}\Omega\left(r\sqrt{\frac{s_1}{s_2}}
\right)^{-1/2}W_{0,m}\left(\beta\sqrt{\frac{s_1}{s_2}}r\right)x_A
\end{equation}
2. $N_{0}^{2}r^{2}s_1/s_2\ll1$
\begin{equation}
{\cal F}_{A}=-4\mu N_{0}^{2}C_{s}\Omega\frac{1}{\sqrt{\gamma}}
\sinh\left(\gamma^{2}\frac{s_1}{s_2}r^{2}\right)x_A
\end{equation}
It is important to note that in the strict sense we have to use
the general expressions for the forces given by the gauge model
(see~\cite{edelen,lagoud}). These calculations, however, are too
cumbersome and will be omitted here. Notice only that the main
conclusions agree very closely with (60)-(62).

\section{Conclusion}

In this paper we have studied the vortex-like solutions for
disclinations within the most general gauge model of rotational
defects when both disclinations and rotational dislocations are
taken into account. The model contains two additional parameters,
coupling constants $s_1$ and $s_2$, which allow us to introduce
two characteristic lengths $l_{\phi}$ and $l_W$. The appearance of
these lengths is the unique property of the gauge model that
distinguishes it from the classical elasticity theory as well as
from other known models of elastic continuum with topological
defects.

There are two distinctive features of the vortices in elastic
media. First, the elastic flux is 'classical' in its origin, i.e.
there is no quantization as opposed to the magnetic vortex. This
means that generally there are no restrictions on the value of
$\nu$ in (22) apart from $\nu>-1$ for topological reasons.
However, if we take into account the symmetry group of the
underlying crystal lattice the available values of $\nu$ become
'quantized' in accordance with this group (e.g., $\nu =1/6, 1/4,
1/3,...$ for hexagonal lattice). Second, the singular character of
the solution on the defect line is typical for the dislocation
theory. As a result, all the known solutions for dislocations
contain a logarithmic divergence in the energy. To avoid this
difficulty, one introduces a cut-off from below by using $r_0$ as
a core radius of the defect. The core region itself is assumed to
be beyond the scope of the linear theory of elasticity. For this
reason, any nonsingular solution will be of essential interest.

Finally, let us note that a similar problem appears for point
$4\pi$ disclinations. In this case, the gauge group $SO(3)$ should
be considered. We expect that the inclusion of rotational
dislocations will clarify the role of the characteristic lengths
as well as the problem of the core region in this case. This study
is now in progress.

\vskip 0.5cm
This work has been supported by the Russian Foundation
for Basic Research under grant No. 97-02-16623, and the Slovak
Scientific Grant Agency, Grant No. 7043.

\end{document}